\documentclass[amssymb,amsfonts,aps,pra,superscriptaddress,onecolumn,showpacs,reprint]{revtex4-1}

\usepackage[bookmarks=false,colorlinks=true, linkcolor=blue,urlcolor=blue,citecolor=blue,pdfstartview={FitH}]{hyperref}

\usepackage{graphicx}% Include figure files
\usepackage{dcolumn}% Align table columns on decimal point
\usepackage{bm}% bold math

\begin{document}

%\preprint{AIP/123-QED}

\title[Sample title]{High-accuracy large-scale DFT calculations using localized orbitals in complex electronic systems: The case of graphene-metal interfaces}

\newcommand{\uamfmc}{Departamento de F\'{i}sica de la Materia Condensada, Universidad Aut\'{o}noma de Madrid, E-28049 Madrid, Spain}
\newcommand{\uamftmc}{Departamento de F\'{i}sica Te\'{o}rica de la Materia Condensada, Universidad Aut\'{o}noma de Madrid, E-28049 Madrid, Spain}
\newcommand{\ifimac}{Condensed Matter Physics Center (IFIMAC), Universidad Aut\'{o}noma de Madrid, E-28049 Madrid, Spain}
\newcommand{\nims}{First-Principles Simulation Group, Nano-Theory Field, International Center for Materials Nanoarchitectonics (WPI-MANA),
National Institute for Materials Science (NIMS), 1-1 Namiki, Tsukuba, Ibaraki 305-0044, Japan}
\newcommand{\ucl}{Department of Physics \& Astronomy, University College London, Gower Street, London WC1E 6BT, U.K.}
\newcommand{\lcn}{London Centre for Nanotechnology, University College London, 17-19 Gordon Street, London WC1H 0AH, U.K.}

\author{Carlos Romero-Mu\~{n}iz} \affiliation{\uamftmc} \affiliation{\nims}

\author{Ayako Nakata} \affiliation{\nims}

\author{Pablo Pou} \affiliation{\uamftmc} \affiliation{\ifimac}

\author{David R. Bowler} \affiliation{\ucl} \affiliation{\lcn} \affiliation{\nims}

\author{Tsuyoshi Miyazaki} \affiliation{\nims}

\author{Rub\'{e}n P\'{e}rez} \affiliation{\uamftmc} \affiliation{\ifimac}

\date{\today}% It is always \today, today,

\begin{abstract}
Over many years, computational simulations based on Density Functional Theory (DFT) have been used extensively to study many different materials at the atomic scale. However, its application is restricted by system size, leaving a number of interesting systems without a high-accuracy quantum description.
In this work, we calculate the electronic and structural properties of a graphene-metal system significantly larger than in previous plane-wave calculations with the same accuracy.
%In this work, we calculate the electronic and structural properties of a large graphene-metal system with high precision, comparable to previous plane-wave calculations, but significantly larger.
For this task we use a localised basis set with the \textsc{Conquest} code, both in their primitive, pseudo-atomic orbital form, and using a recent multi-site approach. This multi-site scheme allows us to maintain accuracy while saving computational time and memory requirements, even in our exemplar complex system of graphene grown on Rh(111) with and without intercalated atomic oxygen. This system offers a rich scenario that will serve as a benchmark, demonstrating that highly accurate simulations in cells with over 3000 atoms are feasible with modest computational resources.

\end{abstract}

% PACS, the Physics and Astronomy Classification Scheme.

\maketitle

%%--------------------------------------------------------------%%
%% INTRODUCTION
%%--------------------------------------------------------------%%
\section{Introduction}
% 1. Need for large-scale calculations
Density Functional Theory (DFT) calculations~\cite{hohenberg1964,kohn1965} have become an essential tool in condensed matter physics, quantum chemistry  and materials science~\cite{martin2004}. Since the seminal application to bulk Si \cite{yin1980,yin1982}, methodological developments have extended the application of this quantum mechanical description to increasingly larger systems including complex surface reconstructions, DNA and protein fragments. However, even using the latest improvements in parallel computing and the most sophisticated numerical methods in self-consistent calculation, there are still many relevant problems that fall beyond the scope of DFT, for example, proteins in their native biological environment, defective crystals, large organic molecules on surfaces, and self-assembled monolayers with large periodicity.

The mechanism of oxygen intercalation into monolayer graphene (G) on Rh(111) surface is another example. We previously reported that a highly corrugated graphene layer grown on Rh(111) can be flattened by the intercalation of oxygen atoms, with careful experimental control of the oxygen dosage and temperature \cite{martinrecio2015,romeromuniz2016}. To study this interesting phenomenon, we performed plane-wave DFT calculations and found that the decoupling of the graphene layer from the substrate takes place when oxygen atoms are intercalated in the lowest moir\'{e} sites, consistent with the experimental observation. However, these simulations were limited in  size; in general, to understand novel phenomena of great technological interest like interface reactions~\cite{ferrighi2016} or edge-related processes~\cite{wong2016}, we need to describe large systems whose treatment is almost impossible by conventional plane-wave DFT methods.

This limitation of current DFT simulations cannot be simply overcome by increasing the computer power. DFT calculations based on a plane-wave basis, despite its mathematical robustness and accuracy, often have a serious problem in parallel efficiency.  Fast Fourier Transforms and very smooth pseudopotentials clearly reduce the computational cost, but they require all-to-all communication between cores and this communication time grows rapidly when the number of processors is increased~\cite{bowler2012}. On the other hand, the use of basis functions with finite extension \cite{sankey1989,artacho1999,junquera2001,ozaki2003,ozaki2004,torralba2008,lewis2011} leads to sparse matrices that are naturally suitable for parallel calculations. This sparsity, and the localization properties of the density matrix, lie behind the linear or $\mathcal{O}(N)$ scaling methodologies~\cite{goedecker1999,bowler2012} that have been implemented in different DFT codes including \textsc{Conquest}~\cite{bowler2002}, ONETEP~\cite{skylaris2005}, BigDFT~\cite{mohr2015}, OpenMX~\cite{ozaki2006} and SIESTA~\cite{ordejon1993,ordejon1995}. With these tools, preliminary simulations have been accomplished on different large-scale systems including, among other, complex electrochemical systems \cite{ohwaki2012}, medium-size enzymes with thousands of atoms \cite{lever2014}, membrane ion channels \cite{todorovic2013} or hydrated DNA fragments \cite{otsuka2008}. Moreover, calculations on millions of atoms have been shown to be possible\cite{bowler2010}.

While the use of local basis functions has a great advantage in efficiency, it is very important to be able to reach accuracy comparable to plane wave calculations. Some local basis sets, like ``blip'' functions in {\sc Conquest} or ``psinc'' functions in ONETEP can achieve plane-wave accuracy systematically, but the total number of basis functions is, in general, large. On the other hand, with the pseudo atomic orbital (PAO) basis sets, the number of basis functions is significantly smaller than plane-wave basis sets and the cost of the calculations is much lower, though it is difficult to achieve systematic convergence. Especially, systems that include different types of bonding interactions are particularly challenging for PAO basis sets.

This is the case for graphene adsorbed on metals~\cite{batzill2012,tetlow2014}. Graphene deforms in order to enhance the bonding with the substrate, resulting in a corrugated structure. The simultaneous presence of strong in-plane $\sigma$-bonds and weaker $\pi$-bonds with delocalized out-of-plane charges makes it difficult to describe the mechanical response. The insertion of oxygen atoms further changes the bonding. To achieve high accuracy with PAO basis sets, we generally need to increase the number of basis functions in each atom, that is the use of multiple-$\zeta$ basis sets, especially for some subtle properties, (i.e. adsorption energies, band gaps, etc.). However, the use of large basis sets results in a significant increase of both computational time and memory requirements, as these scale with the cube and square of the basis size, respectively.

The recently developed multi-site support function (MSSF) method \cite{rayson2009,nakata2014,nakata2015} can overcome this problem of computational cost when accurate multiple-$\zeta$ basis sets are used for large systems. With this method, we construct relatively large radius localized orbitals, called multi-site support functions, for each atom to calculate the Kohn-Sham wave functions or density matrix. Each multi-site (MS) support function is expressed as a linear combination of PAOs of the central atom and its neighbouring atoms, and its coefficients are optimized depending on its local environment. This method is very powerful because it is able to reduce the number of local orbitals to the same size as a minimal basis set, while preserving the accuracy of multiple-$\zeta$ basis set calculation.

In this work, we demonstrate the accuracy and efficiency of the MSSF method in calculations of subtle structural and electronic properties of the G/Rh(111) and G/O/Rh(111) systems (the MSSF method has not been applied to a system of this complexity yet).  We first show that using multiple-$\zeta$ basis sets (DZP for Rh and O, and TZDP for C), we can achieve comparable accuracy to plane-wave calculations for the structural and electronic properties of these systems. We systematically compare our previous accurate plane-wave calculations~\cite{martinrecio2015,romeromuniz2016} with the multiple-$\zeta$ PAO calculations using the \textsc{Conquest} code~\cite{hernandez1996,bowler2006,bowler2010}.
Then, we show that almost the same accuracy can be achieved using the MSSF method, implying that plane-wave
calculations can be reproduced with the MSSF method. Since the computational time of DFT calculations is proportional to the cube of the number of MS support functions, and the number of MS support functions is the same as that of minimal basis set, we can dramatically reduce the cost of DFT calculations.
We demonstrate that, using the MSSF approach, we can perform accurate DFT simulations of a challenging
system like G/Rh(111) containing more than three thousands atoms with relatively modest computational resources. This MSSF method, currently implemented in \textsc{Conquest} but transferable to other localized atomic-orbital DFT codes, has the potential to extend large-scale calculations with thousands of atoms to other technologically relevant systems that include different types of bonding interactions.

The rest of the paper is organized as follows.
We first illustrate the model systems of G/Rh(111) and G/O/Rh(111) in Sec.~\ref{sec:model}.
Then, we explain the MSSF method and the details of the calculations in Sec.~\ref{sec:MSmethod} and ~\ref{sec:CalcCondition}, respectively. In Sec.~\ref{sec:structure}, we provide a detailed comparison of the structural properties of these systems calculated with the present PAO basis sets and our previous plane-wave calculations.
Next, we investigate the accuracy of the MSSF method in Sec.~\ref{sec:MSresult}. Then, in Sec.~\ref{sec:CPUtime}, we report the CPU time and parallel efficiency of the MSSF calculations for large G/Rh(111) systems. Finally, concluding remarks are given in Sec.~\ref{sec:Conclusion}.

\section{Computational methods \\ and basis sets\label{sec:ComputationalMethods}}

\begin{figure*}[t]
\includegraphics[width=0.87\textwidth]{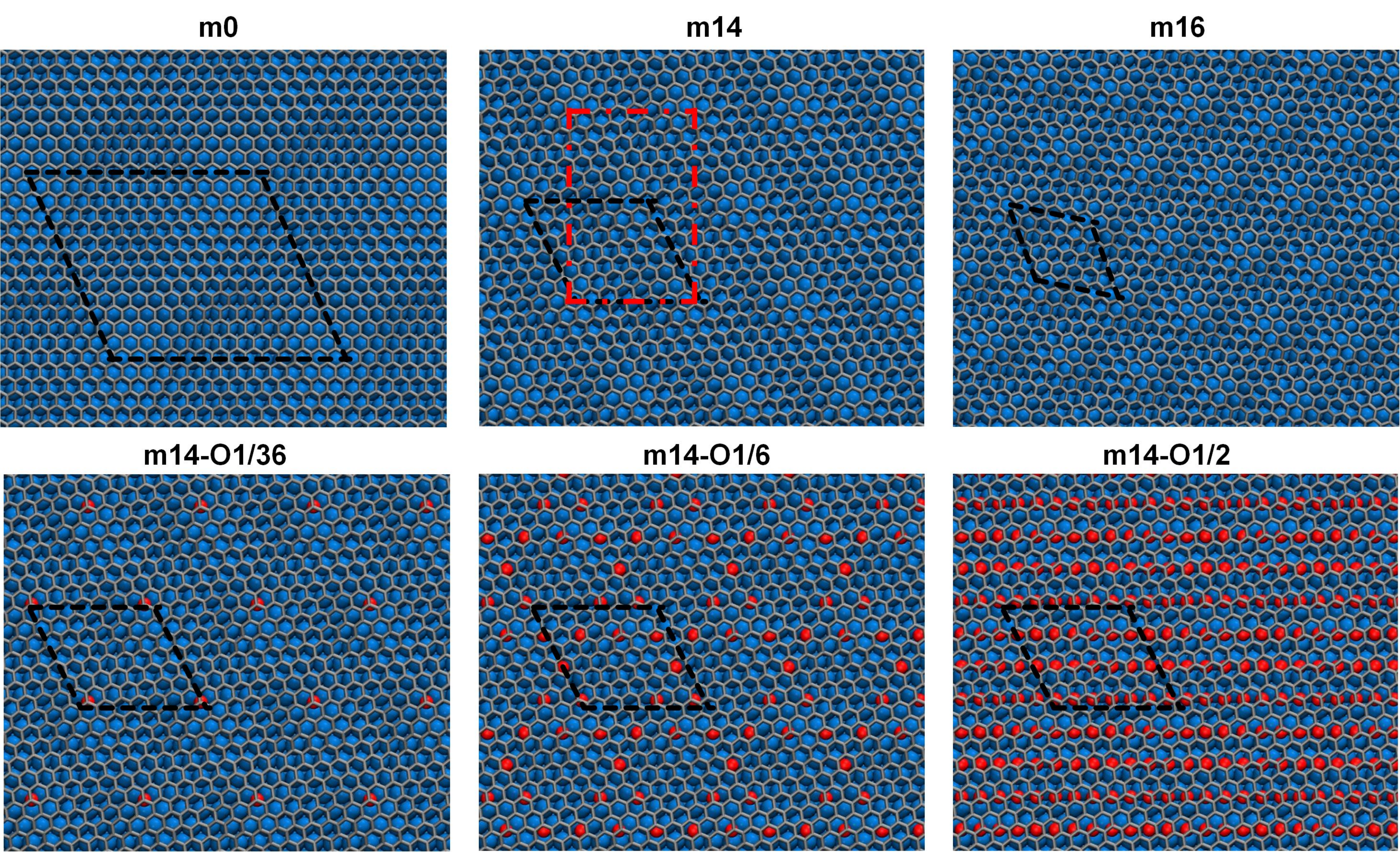}
\caption{\label{structures_fig} Schematic representations (top views) of the six basic moir\'{e} patterns studied in this work. In these ball-and-stick models, the graphene lattice is represented by gray bonds, intercalated oxygen atoms are depicted in red and blue spheres represent rhodium atoms belonging to the metallic substrate. In each case the unit cell of the moir\'{e} patterns is depicted with black dashed lines. More details are given in Table~\ref{celdas_simulacion}. Note that in structure m14 we have also shown a rectangular non-primitive cell whose area is twice of the primitive one. While the primitive cell is a rhombohedron with side $L$ and angle 120$^{\circ}$, the orthogonal cell is a rectangle whose dimensions are $L\times\sqrt{3}L$.}
\end{figure*}

\begin{table*}[t]
\caption{\label{celdas_simulacion}%
Main parameters of the different unit cells used in the calculations. Depending on the code, hexagonal or non-primitive rectangular cells must be used. The size is given in terms of the length of the lattice vectors. The oxygen coverage of the interface is denoted by $\theta_{\rm{O}}.$}
\begin{ruledtabular}
\begin{tabular}{lccccccccccc}
Structure	&	Cell	&	Size (\AA)	&	\multicolumn{2}{c}{Superstructure}	&	Strain (\%)	&	\multicolumn{3}{c}{No. of atoms}		&	$\theta_{\rm{O}}$	&	$k$-points\footnote{These values are for ionic relaxations. In m14 and related structures, when calculating density of states finer grids are used: $11\times11\times1$ for hexagonal cells and $11\times7\times1$ for rectangular cells.}	\\
	&		&		&	Relative to graphene	&	Relative to rhodium	&		&	C 	&	Rh	&	O 	&		&		\\
\colrule
m16	&	Hex.	&	$12.35 \times 12.35$	&	$5\times5$	&	$(\sqrt{21}\times\sqrt{21})$-R$10.9^{\circ}$	&	$+0.13$	&	50	 &	84	&	0	&	0	&	$3\times3\times1$	\\
	&	Rect.	&	$12.40 \times 21.47$	&		&		&	$+0.13$	&	100	&	168	&	0	&	0	&	$3\times2\times1$	 \\
m14	&	Hex.	&	$16.17 \times 16.17$	&	$(\sqrt{43}\times\sqrt{43})$-R$7.6^{\circ}$	&	$6\times6$	&	$-0.04$	&	144	&	 86	&	0	&	0	&	$2\times2\times1$	\\
	&	Rect.	&	$16.23 \times 28.05$	&		&		&	$-0.04$	&	288	&	172	&	0	&	0	&	$2\times1\times1$	 \\
m14O--1/36	&	Hex.	&	$16.17 \times 16.17$	&	$(\sqrt{43}\times\sqrt{43})$-R$7.6^{\circ}$	&	$6\times6$	&	$-0.04$	&	 144	&	86	&	1	&	$1/36$	&	$2\times2\times1$	\\
	&	Rect.	&	$16.23 \times 28.05$	&		&		&	$-0.04$	&	288	&	172	&	2	&	$1/36$	&	$2\times1\times1$	 \\
m14O--1/6	&	Hex.	&	$16.17 \times 16.17$	&	$(\sqrt{43}\times\sqrt{43})$-R$7.6^{\circ}$	&	$6\times6$	&	$-0.04$	&	 144	&	86	&	6	&	$1/6$	&	$2\times2\times1$	\\
	&	Rect.	&	$16.23 \times 28.05$	&		&		&	$-0.04$	&	288	&	172	&	12	&	$1/6$	&	$2\times1\times1$	 \\
m14O--1/2	&	Hex.	&	$16.17 \times 16.17$	&	$(\sqrt{43}\times\sqrt{43})$-R$7.6^{\circ}$	&	$6\times6$	&	$-0.04$	&	 144	&	86	&	18	&	$1/2$	&	$2\times2\times1$	\\
	&	Rect.	&	$16.23 \times 28.05$	&		&		&	$-0.04$	&	288	&	172	&	36	&	$1/2$	&	$2\times1\times1$	 \\
m0	&	Hex.	&	$29.66 \times 29.66$	&	$12\times12$	&	$11\times11$	&	$+0.15$	&	288	&	484	&	0	&	0	&	 Gamma	\\
	&	Rect.	&	$29.75 \times 51.53$	&		&		&	$+0.15$	&	576	&	968	&	0	&	0	&	Gamma	\\
m0double	&	Rect.	&	$59.51 \times 51.53$ & $24\times12$	&	$22\times11$	 	&	$+0.15$	&	1152	& 1936	& 0 &	0	 &	 Gamma	 \\
\end{tabular}
\end{ruledtabular}
\end{table*}

%\subsection{Systems and simulation cells}
\subsection{Simulation cells for the G/Rh moir\'{e}s and O intercalation.\label{sec:model}}

The first system that we study in this work is the G/Rh(111) interface. Although this interface is regarded as a strongly interacting system, it displays a number of different rotational domains or moir\'{e} patterns as a consequence of the subtle balance between the corrugation and interaction contributions to the total energy~\cite{martinrecio2015}. Graphene adopts a rippled structure (corrugations about 1\,{\AA}) highly hybridized with the substrate and with strong modulations in the adsorption distance from the Rh (111) surface~\cite{martinrecio2015}. This highly-coupled state dramatically affects the electronic properties of graphene, leading to the complete destruction of the characteristic Dirac cones~\cite{batzill2012,tetlow2014}.

However, the linear dispersion of pristine graphene can be recovered by intercalating oxygen atoms at the interface. We also work on this G/O/Rh(111) system, with various amount of oxygen atoms. The intercalation of oxygen atoms leads to a step--by--step decoupling of the G layer, where the G corrugation and the electronic properties vary depending on the amount of intercalated oxygen atoms, evolving from a purely chemisorbed state to a quasi-free-standing flat monolayer, physisorbed to the substrate by dispersion forces. Finally, when an ordered O-$(2\times1)$ network of atomic oxygen is formed at the interface, the corrugation almost disappears and the Dirac cones are restored except for a small energy shift due to the charge transfer~\cite{romeromuniz2016}.

For G/Rh(111) in this study, we consider three of these different moir\'{e} patterns that are experimentally observed: m0, m14 and m16, using the nomenclature established in ref.~\citenum{martinrecio2015}. The structure m0 $(12\times12)_{\rm{G}}$ is the one identified in earlier studies and the most frequently observed, while m14 and m16 are two smaller moir\'{e}s that spanned the range of moir\'{e} sizes identified in our experiments~\cite{martinrecio2015}.
They are modeled using unit cells of different sizes created by a single layer of graphene on top of a four-layer rhodium slab with a vacuum region whose thickness along the $z$ axis is larger than 14 {\AA}.
The details of the different unit cells used are summarized in Table~\ref{celdas_simulacion} and depicted with ball--and--stick models in Fig.~\ref{structures_fig}.

There are two kinds of unit cells presented in Table~\ref{celdas_simulacion}. The $L \times L$ primitive rhombohedral unit cells were used in our previous calculations in ref.~\citenum{martinrecio2015}. We used the VASP (Vienna Ab initio Simulation Package) code \cite{kresse1996} for these plane-wave DFT calculations. Since only rectangular unit cells are available in the current version of \textsc{Conquest}, $L \times \sqrt{3}L$ rectangular unit cells, which are larger but equivalent to the rhombohedral unit cells, are used in the \textsc{Conquest} calculations. The rhombohedral and rectangular unit cells are shown as the black and red boxes in Fig.~\ref{structures_fig}.

It is important to note that the procedure used to build these cells in order to adjust the mismatch between the two lattices is the same used in ref.~\citenum{martinrecio2015}. We assume that the graphene layer has to deform to match the unaltered metallic substrate. In this way, the experimental strain of the graphene layer is preserved in our calculations, where we use the equilibrium lattice parameters obtained in the simulations with each different method (VASP or \textsc{Conquest}).
Since lattice parameters for G and bulk Rh are slightly different depending on the method, the sizes of the same unit cell quoted in Table~\ref{celdas_simulacion} are also different.

% O intercalation
The intercalation of oxygen is modeled by introducing a variable number of oxygen atoms at the G/metal interface in the m14 moir\'{e}, that shares the main features of the m0 moir\'{e} but allows faster calculations. Based on our previous study~\cite{romeromuniz2016}, we have considered three different oxygen coverages, that expanded the whole decoupling process, from the accumulation of oxygen on the high moir\'{e} areas in the early stages (with coverages of 1 and 6 O atoms per unit cell, in the m14--O1/36 and m14--O1/6 cases), to the formation of an ordered O$-(2\times1)$ network of atomic oxygen at the interface, that leads to a complete decoupling~\cite{romeromuniz2016}.

\subsection{Local orbital basis sets and multi-site support functions\label{sec:MSmethod}}

In contrast to the plane-wave basis functions used in our previous calculations~\cite{martinrecio2015,romeromuniz2016}, real-space local orbitals, called support functions, are used in \textsc{Conquest} to express the Kohn-Sham orbitals and density matrix \cite{hernandez1996,bowler1999,bowler2006,bowler2010,sena2011,arita2014}. Support functions are constructed to be localized, i.e., to vanish beyond a finite range. This locality enables us to reduce the computational cost significantly. In \textsc{Conquest}, the support function, $\phi_{i\alpha}(\mathbf{r})$, is built as a linear combination of some localized basis functions $\xi_{i\mu}(\mathbf{r})$, associated with each atom $i$ as
\begin{equation}\label{singleorbital}
\phi_{i\alpha}(\mathbf{r}) = \sum_{\mu}b_{i\alpha,i\mu}\xi_{i\mu}(\mathbf{r}),
\end{equation}
with $b_{i\alpha,i\mu}$ is the linear-combination coefficient.

\begin{table}[t]
\caption{\label{parameters}%
Lattice parameters and bulk moduli obtained with DFT following the different methodologies.}
\begin{ruledtabular}
\begin{tabular}{lccc}
 & Graphene      & \multicolumn{2}{c}{Bulk rhodium} \\
 & $a_0$ ({\AA}) & $a_0$ ({\AA})  &  $B_0$ (GPa) \\
\colrule
VASP        &   2.4678 &  3.7729  &  270 \\
\textsc{Conquest}    &   2.4762 &  3.7903  &  255 \\
\textsc{Conquest}-MSSF &   2.4767 &  3.7844  &  265 \\
Exp.        &   2.46   &   3.80   &  269 \\
\end{tabular}
\end{ruledtabular}
\end{table}

\begin{figure}[t]
\includegraphics[width=0.5\textwidth]{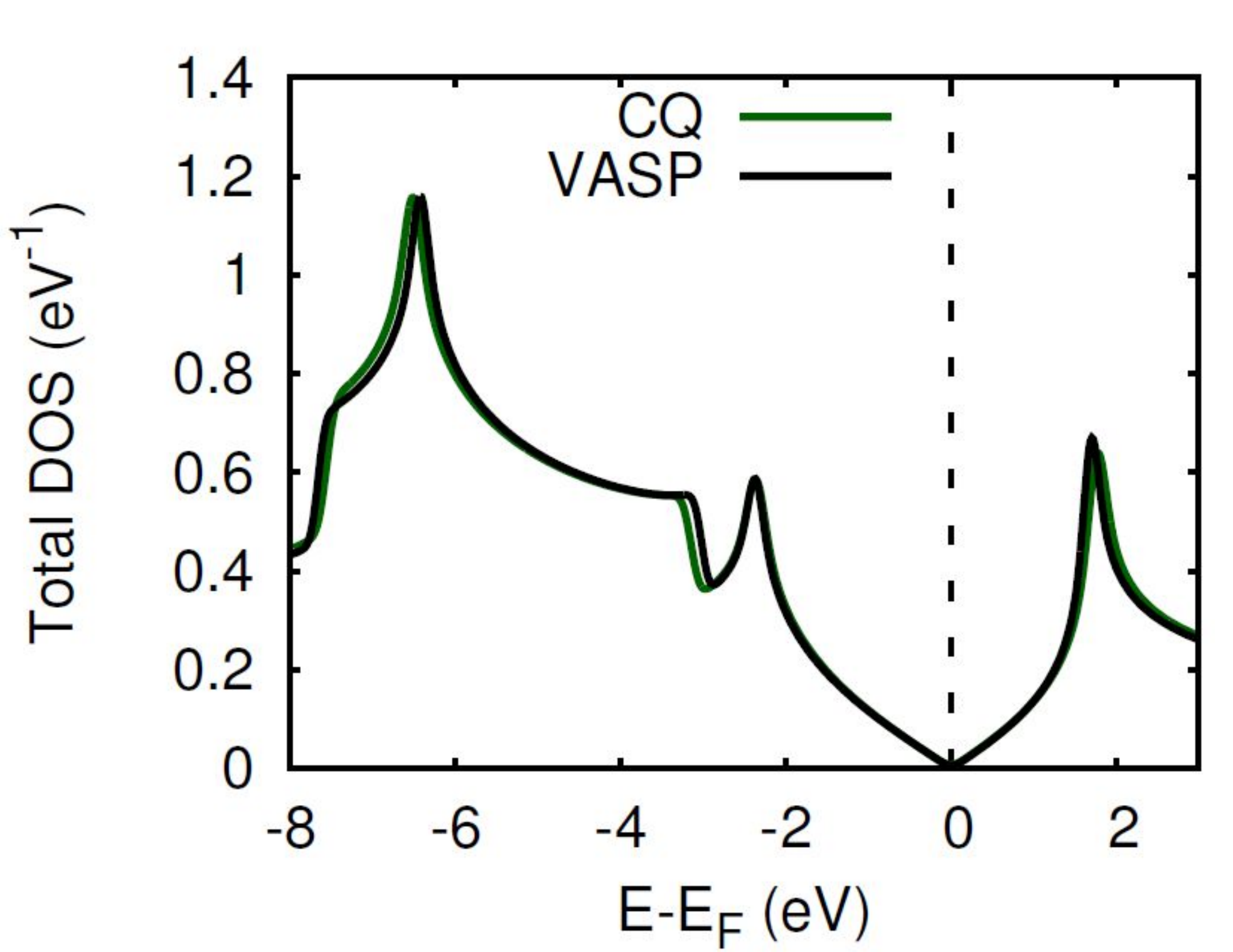}
\caption{\label{DOS_graphene} Total density of states for free-standing graphene calculated with plane waves (black solid line) and localized orbitals (green solid line).}
\end{figure}

\textsc{Conquest} supports two kinds of localized basis functions $\xi_{i\mu}(\mathbf{r})$, PAOs~\cite{torralba2008} and B-splines on regular grids~\cite{hernandez1997}. The latter basis functions (called ``blip'' functions) are akin to plane-wave basis functions, which can be improved systematically by making the regular grids finer but usually require much longer computation times than PAOs. Therefore, we use PAOs in this work. PAOs consist of the product of a numerical radial function (different functions are used for the same angular momentum, often referred to as multiple-$\zeta$) and analytical spherical harmonic functions (s, p, d, $\cdots$). PAOs are used with the pseudopotential to express the electronic configurations of valence electrons. We used double-$\zeta$ and polarization (DZP) PAOs for rhodium, (2 $\times$ s, 2 $\times$ d, 1 $\times$ p) (15 functions in total) with the ranges \{(6.8, 6.6), (4.2, 2.8) and (6.8)\} bohr, and triple-$\zeta$ and triple-polarization (TZTP) PAOs for oxygen, (3 $\times$ s, 3 $\times$ p, 3 $\times$ d) (27 functions in total) with the ranges (6.0, 4.0, 2.0) bohr for all angular momentum. Triple-$\zeta$ and double-polarization PAOs, (3~$\times$ s, 3 $\times$ p, 2 $\times$ d) (22 functions in total) with the ranges \{(6.1, 4.0, 3.0), (7.1, 5.0, 3.0) and (7.1, 5.0)\} bohr, were used for carbon.
When a PAO basis set is used, each support function can be represented by one PAO without any modification (in this case, $b_{i\alpha,i\mu}= 1$ for $\alpha = \mu$, otherwise $0$), or we can optimize the linear-combination coefficient to minimize the total energy of the system.

The accuracy of PAOs of carbon is particularly important to describe the electronic and structural properties of G/Rh (111) systems in this work. The final corrugation of the layer is the result of a subtle interplay between interaction and deformation energies. We first checked the accuracy of the PAOs by calculating the lattice parameters of graphene and rhodium. In Table~\ref{parameters}, we can see that the lattice parameters calculated with the PAOs by \textsc{Conquest} are very close to those with the plane waves by VASP, and to the experimental values. For the electronic properties, we compare in Fig.~\ref{DOS_graphene} the total density of states (DOS) of graphene obtained with plane waves and PAOs. The agreement of the electronic structure, including the linear dispersion around the Fermi energy and the van Hove singularities, is almost perfect. This result demonstrates that the present PAOs of carbon can describe the electronic structure of the undistorted planar graphene with the plane-wave accuracy.

% General description
In order to achieve large-scale DFT calculations, reducing the number of support functions is important, because the computational cost scales with the cube of the number of support functions. The primitive PAO basis sets can be contracted into small number of support functions. Conventionally, as in Eq.~(\ref{singleorbital}), the contraction is taken over the PAOs on each atom, though the number of functions that can be contracted is limited by symmetry\cite{torralba2008}. On the other hand, we have recently introduced an efficient MSSF method to contract PAOs from multiple atoms to the minimal, i.e., single-($\zeta$), size~\cite{nakata2014, nakata2015} while retaining the accuracy of the full basis. In the MSSF method, we take the linear combination of the PAOs not only on the target atom but also on the neighboring atoms within a cutoff region.
In mathematical terms, this means that Eq.~(\ref{singleorbital}) now becomes
\begin{equation}\label{singleorbital_ms}
\phi_{i\alpha}(\mathbf{r}) = \sum_{l}^{\rm{neighbors}}\sum_{\mu\in l}b_{i\alpha,l\mu}\xi_{l\mu}(\mathbf{r}),
\end{equation}
where the summation on $l$ runs over neighboring atoms which are within the cutoff region from the target atom $i$, including atom $i$ itself.
The coefficients $b_{i\alpha,i\mu}$ can be determined by numerical optimization~\cite{nakata2015}, but in this study we determined them by using the local filter diagonalization method~\cite{rayson2009,rayson2010,nakata2014}, in which the coefficients are determined by solving for localized molecular orbitals in a subspace around the target atom.
In our calculations, 15, 27 and 22 PAOs of rhodium, oxygen and carbon atoms are contracted into 6, 4 and 4 multi-site support functions , respectively. We set the radii of the cutoff region of the MSSFs and the subspace for the localized molecular orbitals both to 16 bohr. Note the good agreement achieved with the MSSF method in the equilibrium parameters of graphene and rhodium collected in Table~\ref{parameters}.
These results demonstrate that the MSSF method reproduces the accuracy of the full basis set, while reducing the size of the support space.

It is worth pointing out that the MSSF method, especially when the MSSFs are calculated by numerical optimization, may not have good accuracy for the unoccupied states since MSSFs are optimized only for the accurate description of occupied Kohn-Sham orbitals or density matrix. But, even in such cases, the calculated ground state charge density obtained by the MSSF method is accurate and we can evaluate the Hamiltonian matrix with primitive PAO basis set using the charge density. Then, we can calculate the eigenvalues of the unoccupied states accurately~\cite{nakata2017}, by using an efficient algorithm~\cite{sakurai2003} that provides access to the eigenstates in a specific energy window.

\subsection{Simulation details\label{sec:CalcCondition}}

As in our previous plane-wave DFT simulations, we use a generalized gradient approximation for the exchange and correlation functional as described by Perdew, Burke and Ernzerhof (PBE)~\cite{perdew1996}, with the D2 semi-empirical correction~\cite{grimme2006} to take into account dispersion forces which are essential in the proper determination of the corrugation values~\cite{wang2010,stradi2011}.
A norm-conserving pseudopotential is used in \textsc{Conquest}, while Projector-Augmented Wave (PAW) method \cite{bloch1990,kresse1999} is used in VASP.
Electronic self-consistency is achieved with a tolerance of $\sim3\times10^{-6}$ eV. Spin polarization is not required in our calculations because none of the systems presents magnetic behavior. Reciprocal space is sampled using different Monkhorst-Pack grids~\cite{monkhorst1976} according to the size of each unit cell, as indicated in Table~\ref{celdas_simulacion}. In the calculations of the density of states in Sec.~\ref{sec:MSresult}, the number of $k$-points was increased to gain accuracy.

The geometry optimizations were performed using the conjugate gradient algorithm. During these optimizations, the two bottom layers of the slab were kept fixed in their bulk positions while the rest, including oxygen atoms, were allowed to relax.
The geometry optimizations with \textsc{Conquest} were started from the geometries pre-converged by VASP.
The geometry optimization using the conjugate gradient method was performed until the energy change becomes smaller than $10^{-6}$ eV/atom or the maximum force is smaller than 0.05 eV/{\AA}.
In adsorption energy calculations, the counterpoise method was used to correct the basis set superposition error (BSSE)~\cite{duijneveldt1994} which comes from the use of the localized PAOs.

\section{Results and discussion}

\subsection{Structural properties and energetics\label{sec:structure}}
% TM: First, structure of G/Rh(111)
In this section, we investigate the accuracy of the primitive PAO basis sets for the structural properties and energetics in the G/Rh(111) and G/O/Rh(111) systems, by comparing the present PAO calculations with \textsc{Conquest} and our previous plane-wave calculations with VASP.

For G/Rh(111) system, we compare the optimized geometries of m16, m14 and m0 moir\'{e} patterns shown in Table~\ref{celdas_simulacion} and Fig.~\ref{structures_fig}.
Table~\ref{structures_nooxy} shows the minimum and maximum heights for the carbon atoms in the G layer and the corrugation, which is defined as the difference between minimum and maximum heights. In all cases, structure optimizations are performed with the initial geometry provided by VASP. We also performed the geometry optimization of the m14 structure starting from a flat geometry in the graphene layer, as we previously did in the VASP calculation, and confirmed that no other local minimum but the corresponding corrugated structure was found by the geometry optimization.
The structural parameters in Table~\ref{structures_nooxy} show good agreements between \textsc{Conquest} and VASP results, and the differences are smaller than 3\%. The only significant discrepancy appears in the larger moir\'{e} (m0), where a difference in the maximum height of 0.10~\AA\ (3.05~\AA\ for \textsc{Conquest} versus 3.15~\AA\ for VASP) leads to a corrugation of 1.12~\AA\ in \textsc{Conquest} while 1.21~\AA\ obtained with VASP.
However, it should be noted that, due to the large size of this structure, its corrugation value is very sensitive to strain conditions or the computational description of the system. For instance, we can find in the literature~\cite{iannuzzi2011,voloshina2012} other values calculated with VASP for slightly different strain conditions as shown in Table \ref{structures_nooxy}. Therefore, the discrepancies obtained between both codes for this structure are still in agreement within the expected uncertainties.
We also note that the corrugation of the graphene layer in this kind of systems is very subtle, and minimal variations in the description of the interaction lead to very different values.
For example, we have checked that the use of one of the recently proposed non-local vdW exchange and correlation functionals, optB86b \cite{klimes2010,klimes2011}, instead of the PBE+D2 using the VASP code leads to differences in corrugations as large as $10\%$.

More importantly, experiments show that these three moir\'{e} patterns display linear growth of the graphene corrugation as a function of the moir\'{e} unit cell size~\cite{martinrecio2015}. Note that this trend is quantitatively reproduced with the PAO basis (see Table~\ref{structures_nooxy}).

\begin{table}[t]
\caption{\label{structures_nooxy}%
Main structural parameters of the equilibrium structures for different moir\'e patterns. $z_{min}$ and $z_{max}$ are the minimum and maximum heights of the graphene layer with respect to the rhodium surface. $C_G$ is the total corrugation of the layer.}
\begin{ruledtabular}
\begin{tabular}{lcccc}
Structure	&	Code	&	$z_{min}$	&	$z_{max}$	&	$C_{\rm{G}}$	 \\
        	&	     	&	({\AA})	    &	({\AA})   	&	({\AA})	 \\
\colrule
m16	        &	VASP	&	2.01	&	2.94	&	0.92	\\
	        &	CQ	    &	2.01	&	2.92	&	0.90    \\
m14	        &	VASP	&	2.07	&	3.14	&	1.07	\\
	        &	CQ	    &   2.04    &	3.08	&	1.04    \\
m0	        &	VASP	&	1.94	&	3.15	&	1.21	\\
	        &	CQ	    &	1.93	&	3.05	&	1.12	\\
	        &	VASP\footnote{From Ref. \cite{voloshina2012}.}	&	2.08	&	3.15	&	1.07	\\
\end{tabular}
\end{ruledtabular}
\end{table}

% evolution under O intercalation
We now turn to the G/O/Rh(111) system, for the study of the structural changes in the corrugation and adsorption distance by the intercalation of oxygen atoms. We use only one of the rotational domains (m14) in this analysis. Table~\ref{structures_oxy} shows the structural parameters of m14 with the intercalating oxygen. As we saw with the results shown in Table~\ref{structures_nooxy}, PAO calculations by \textsc{Conquest} can again reproduce the plane-wave results quantitatively. By comparing Table \ref{structures_nooxy} and Table~\ref{structures_oxy}, it is found that the initial corrugation of graphene is increased by 0.28~\AA\, both in \textsc{Conquest} and VASP, in the first stages of intercalation (m14--O1/36). This increase is due to the fact that the oxygen atoms prefer to stay under the highest parts of the moir\'{e} pattern, increasing the value of the maximum height, $z_{max}$. Then, for an intermediate coverage (m14--O1/6), the corrugation decreases when the oxygen atoms begin to occupy the lowest areas, and the decoupling of the substrate starts to take place, as reflected by the dramatic increase in the minimum height ($z_{min}$). Corrugation values for VASP (0.80~\AA) and \textsc{Conquest} (0.84~\AA) for this particularly complex stage in the process, where both the minimum and maximum heights are rapidly changing, match to within 5\%. Finally, for higher oxygen coverage (m14--O1/2), the decoupling is achieved and the corrugation practically disappears, leading to the same value (0.11~\AA) in both methods. These results show that the present PAO basis sets are extremely accurate in different environments, ranging from a highly-coupled chemisorbed state to a fully-decoupled physisorbed regime with an adsorption distance of $\sim 3.5$ {\AA} characteristic of binding by dispersion forces. Thus, the PAO basis is able to reproduce the evolution of the graphene-metal interaction as accurately as a plane-wave basis.

\begin{table}[t]
\caption{\label{structures_oxy}%
Main structural parameters of the equilibrium structures for different oxygen coverages. $z_{min}$ and $z_{max}$ are the minimum and maximum heights of the graphene layer with respect to the rhodium surface. $C_G$ is the total corrugation of the layer.}
\begin{ruledtabular}
\begin{tabular}{lcccc}
Structure	&	Code	&	$z_{min}$	&	$z_{max}$	&	$C_{\rm{G}}$	 \\
        	&	     	&	({\AA})	    &	({\AA})   	&	({\AA})	 \\
\colrule
m14O--1/36	&	VASP	&	2.01	&	3.36	&	1.35	\\
	        &	CQ	    &  	2.03    &	3.35	&	1.32	 \\
m14O--1/6	&	VASP	&	2.83	&	3.62	&	0.80     \\
	        &	CQ	    &	2.73	&	3.59	&	0.84	 \\
m14O--1/2	&	VASP	&	3.87	&	3.98	&	0.11	\\
	        &	CQ	    &	3.84	&	3.95	&	0.11    \\
\end{tabular}
\end{ruledtabular}
\end{table}

% Energetics
We now consider the energetics of G/Rh(111) systems, which is related to the formation of the different moir\'{e} patterns. Table~\ref{energies} compares the adsorption energy $E_{\rm{ad}}$ and interaction energy $E_{\rm{int}}$ of the G/Rh(111) system for the m16 and m14 structures calculated by VASP with the plane waves and those by \textsc{Conquest} with the PAOs.
Here, the adsorption energy, $E_{\rm{ad}}$, is defined as:
\begin{equation}
E_{\rm{ad}} = E[\rm{G/Rh(111)}] - E[\rm{G}] - E[\rm{Rh(111)}],
\end{equation}
where $E[\rm{G}]$ and $E[\rm{Rh(111)}]$ are the energies of the isolated graphene layer and the metallic slab respectively.
On the other hand, we define the interaction energy $E_{\rm{int}}$ as:
\begin{equation}
E_{\rm{int}} = E[\rm{G/Rh(111)}] - E[\rm{G}]' - E[\rm{Rh(111)}]',
\end{equation}
where $E[\rm{G}]'$ and $E[\rm{Rh(111)}]'$ represent the energies of G and Rh(111) at the equilibrium geometry in the moir\'{e} pattern. By introducing the distortion energies of the subsystems as $\Delta E(\rm{G})=E[\rm{G}]'-E[\rm{G}]$ for graphene and $\Delta E(\rm{Rh})=E[\rm{Rh(111)}]'-E[\rm{Rh(111)}]$ for rhodium, we can express the adsorption energy as
\begin{equation}
E_{\rm{ad}} = E_{\rm{int}} + \Delta E(\rm{G}) + \Delta E(Rh).
\end{equation}
Note that the distortion energies include two contributions: a small strain contribution, and the predominant term related to the corrugation of both subsystems ~\cite{martinrecio2015}. While the adsorption energy is related to the stability of the different moir\'e structures, the interaction energy quantifies the gain associated with the G-metal interaction either by the creation of C-metal bonds or merely by weak dispersive interactions in physisorbed systems.

From Table~\ref{energies}, we can see quantitative agreement between \textsc{Conquest} and VASP results, for both m14 and m16 structures.
The difference of the interaction energy is 9 meV and 16 meV for m14 and m16 structures, respectively. The difference in the distortion energy is 2 meV and 4 meV for graphene, and 6 meV and 1 meV for rhodium. As a result, the difference of adsorption energy is about 10 meV for both m14 and m16 structures. We might be able to reduce this difference between VASP and \textsc{Conquest}, if we used a larger basis set for Rh; but the differences are already much smaller than the absolute values of the interaction, distortion and adsorption energies.  These differences are close to the limit for the agreement between different DFT codes: in particular, \textsc{Conquest} uses norm-conserving pseudopotentials while VASP uses the PAW method.
In addition, if we compare the relative stability of the two moir\'{e} structures, the energy difference is even smaller; m16 is more stable than m14 structure by 4 meV and 7 meV for VASP and CONQUEST, respectively.

Among the graphene-metal systems, G/Rh(111) system belongs to the strongly interacting group~\cite{batzill2012}. As can be seen in Table~\ref{energies}, the interaction energy of G/Rh(111) is larger than 170 meV/C atom. Its adsorption energy is also larger than 100 meV/C atom, much larger than in other substrates like Pt(111) ($\sim40$ meV/C atom)\cite{gao2011}.
Due to this strong interaction, it was initially assumed that only one preferential moir\'{e} structure was stable. However, there is a subtle balance between the energetic cost of graphene corrugation and the energy gain associated with the creation of C-metal bonds, which leads to the formation of multiple moir\'{e} patterns with very different sizes but similar adsorption energies~\cite{martinrecio2015}.
We can see that PAO calculations by \textsc{Conquest} nicely reproduce these key aspects of the system, as in the VASP calculations.

\begin{table}[t]
\caption{\label{energies}%
Deformation energies $\Delta E$ for graphene and rhodium, adsorption $E_{\rm{ad}}$ and interaction $E_{\rm{int}}$ energies calculated by VASP, \textsc{Conquest} with PAOs (CQ) and \textsc{Conquest} with MSSFs. Results denoted with the asterisk do not include the BSSE correction, to enable comparison between PAO and MSSF calculations.}
\begin{ruledtabular}
\begin{tabular}{lrrrrrr}
 & \multicolumn{4}{c}{m14} & \multicolumn{2}{c}{m16} \\
E meV/(C atom) & VASP & CQ & CQ$^{*}$ & MSSF$^{*}$ & VASP & CQ \\
\hline
$\Delta E$(G)     &   $24$    &  $22$     &  $22$   &   $22$  &  $40$   &  $36$    \\
$\Delta E$(Rh)    &   $28$    &  $34$     &  $34$   &   $44$  &  $33$   &  $32$    \\
$E_{\rm{int}}$    &  $-180$   &  $-171$   &  $-351$ &  $-367$ &  $-205$ & $-189$  \\
$E_{\rm{ad}}$     &  $-128$   &  $-114$   &  $-295$ &  $-301$ &  $-132$ & $-121$  \\
\end{tabular}
\end{ruledtabular}
\end{table}

%%% Multi-site results for structure and energetics
\subsection{Accuracy of the multi-site support functions.\label{sec:MSresult}}
In this section, we study the accuracy of the MSSF method for the structural, energetic and electronic properties of the G/Rh(111) and G/O/Rh(111) systems. In the previous section, we have confirmed the high accuracy of the primitive multiple-$\zeta$ PAO basis set for the structural and energetic properties of the graphene-metal systems. We now examine the accuracy of the MSSF method by investigating whether  calculations with MSSFs can reproduce the calculations with primitive PAO basis sets, reported in the last section.

For the structural properties, we calculated the forces and the energies of the G/Rh(111) systems with the primitive PAOs and MSSFs using the same geometries obtained by the primitive PAOs. It is found that the differences are very small, below 0.1 eV/{\AA} for the forces and 0.05 eV per atom for energy.
For the energetics, shown in Table~\ref{energies}, we first compare the distortion energy and find good agreement in the case of graphene: 22 meV/C atom for both PAOs and MSSFs. For the rhodium slab, the result is not as good: 34 vs. 44 meV/C atom using primitive PAOs and MSSFs, respectively. It is not clear why we have larger differences for Rh, but they are still acceptable for the present purpose. We expect that the differences would become smaller if we optimized the MSSFs. For the adsorption and interaction energies, since the BSSE correction has not yet been implemented for MSSF in \textsc{Conquest}, we compare the energies without the BSSE correction. The comparison shows that the differences are small, 351 vs. 367 meV/C atom $(\sim7\%)$ for the interaction energies and 295 vs. 301 meV/C atom $(\sim2\%)$ for the adsorption energies.
All of these results support the high accuracy of the MSSF method for the structural and energetic properties of G/Rh(111) system.

\begin{figure*}[t]
\includegraphics[width=0.98\textwidth]{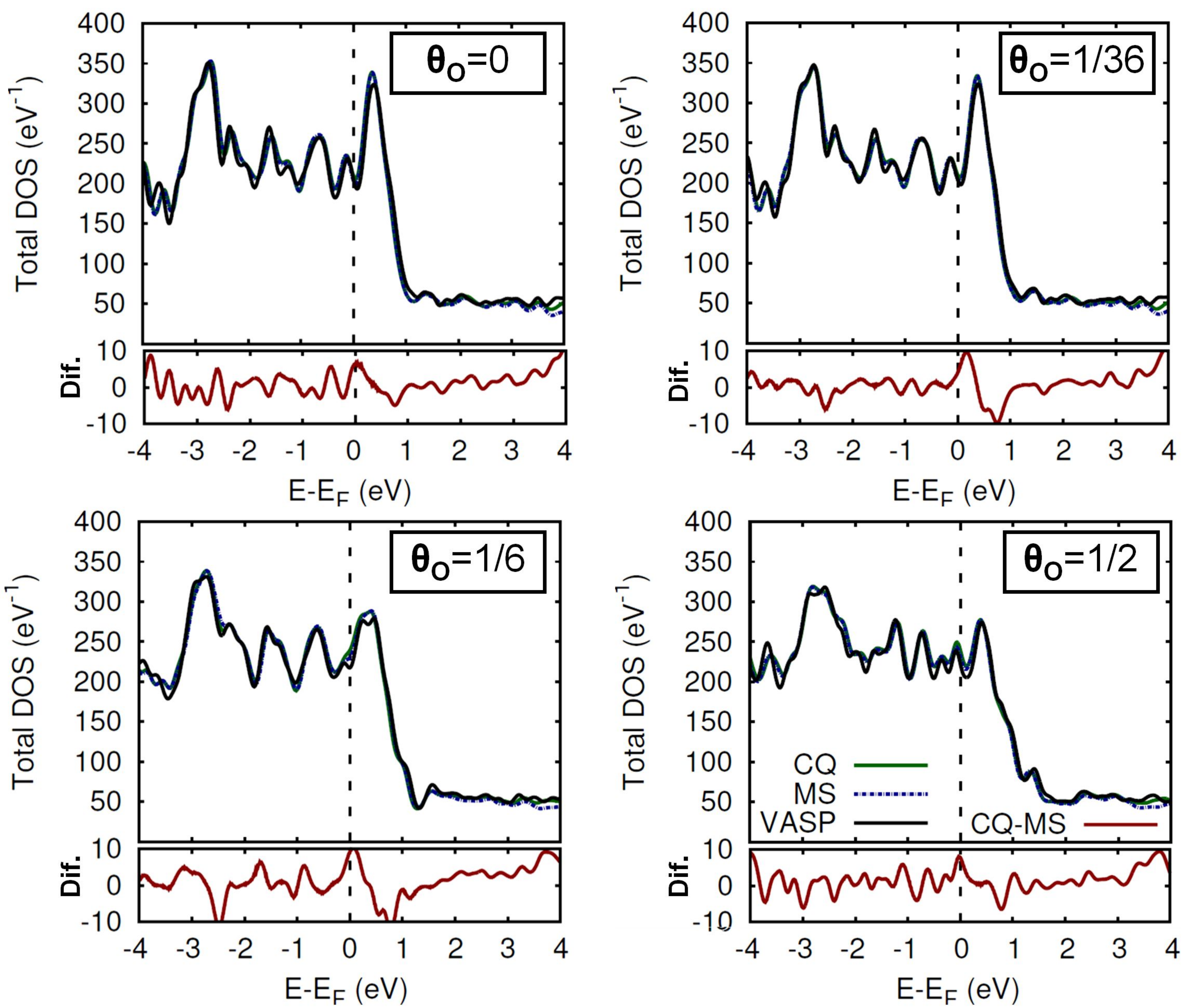}
\caption{\label{DOS} Total density of states for the m14 structure with different oxygen coverage in the interface calculated with plane waves (black solid lines), localized orbitals (green solid lines) and the contracted multi-site orbitals (blue dashed lines). The red lines in the lower panels represent the DOS difference between the standard PAOs of \textsc{Conquest} and the MSSF calculations. As explained in the text, the amount of intercalated oxygen clearly modifies the coupling between the graphene and the metallic substrate, displaying a distinctive behavior in each case.}
\end{figure*}

We next explore the electronic properties of the G/O/Rh(111) system.
Note that the system is appropriate for this task because of the change in electronic coupling of the graphene layer with the amount of intercalated oxygen at the interface.
We first examine the difference between the plane-wave results from VASP and those from the primitive PAO basis sets. Upper panels of Fig.~\ref{DOS}(a)-(d) shows the total DOS by plane-wave and primitive PAO basis sets for different oxygen coverage at the interface, $\theta_{\rm{O}}$. We can see that the DOS near the large peaks located at $E-E_F\sim-3$ eV and $E-E_F\sim+0.5$ eV changes clearly when $\theta_{\rm{O}}$ increases. In all cases, the agreement between the VASP and \textsc{Conquest} results is remarkable.

Then, we switch to the comparison between primitive PAO and MSSF calculations, using the same geometry obtained with the primitive PAOs of \textsc{Conquest}. The difference between the two methods are shown in the lower panels of Fig.~\ref{DOS} (a)-(d), for different $\theta_{\rm{O}}$. Compared these differences with the value of total DOS, the agreement between the primitive PAOs and the MSSFs is almost perfect. From these results, we can conclude that MSSFs can capture almost the same information that is contained in the large, primitive PAO basis set calculations of G/Rh(111) or G/O/Rh(111) systems.

\begin{table*}[t]
\caption{\label{computime}%
Computational times of one self-consistent-field (SCF) step with the PAOs and the MSSFs for the m0 (1544 atoms) and the m0double (3088 atoms) systems. Note that the matrix construction time includes the time for MSSF construction in the case of MSSF.  The ratios of times between different calculations, giving the relative speed-up, are also shown.}
\begin{ruledtabular}
\begin{tabular}{lccccccc}
                                & \multicolumn{2}{c}{m0}   & & \multicolumn{3}{c}{m0double}                           \\ \cline{2-3} \cline{5-8}
Calculation ID                  & \verb|#1| & \verb|#2|    & & \verb|#3| & \verb|#4|    & \verb|#5|    & \verb|#6|    \\
Total no. of atoms              &    1544   &  1544        & &   3088    &   3088       &   3088       &   3088       \\
Function type                   &    PAO    &  MSSF        & &   PAO     &  MSSF        &  MSSF        &  MSSF        \\
No. of basis elements           &   27192   &  8122        & &  54384    &  16244       &  16244       &  16244       \\
No. of MPI cores                &     432   &   432        & &    108    &    108       &    432       &    864       \\
No. of nodes                    &      36   &    36        & &     72    &     72       &     36       &     36       \\
                                &           &              & &           &              &              &              \\
Time per iteration [sec.]       &           &              & &           &              &              &              \\
~Matrix construction (a)        &     64.3  &  400.4       & &  155.7    & 1455.4       &  730.1       &  405.9       \\
%~MSSF construction\footnotemark[1] &   --      &  283.8       & &  --       & 1136.2       &  507.1       &  282.3       \\
~Diagonalization (b)            &   1192.5  &   39.2       & & 37647.7   &  700.8       &  238.0       &  165.9       \\
~Sum of (a) + (b)               &   1256.9  &  439.6       & & 37803.5   & 2156.3       &  968.1       &  571.8       \\
                                &           &              & &           &              &              &              \\
Relative speed-up    &           & \verb|#1/#2| & &           & \verb|#3/#4| & \verb|#5/#2| & \verb|#5/#6| \\
~Matrix construction (a)        &   --      &   0.2        & &    --     &    0.1       &    1.8       &    1.8       \\
%~MSSF construction\footnotemark[1] &   --      &   --         & &    --     &    --        &    1.8       &    1.8       \\
~Diagonalization (b)            &   --      &  33.8        & &    --     &   54.5       &    6.7       &    1.4       \\
~Sum of (a) + (b)               &   --      &   2.9        & &    --     &   17.8       &    2.2       &    1.7       \\
\end{tabular}
\end{ruledtabular}
%\footnotetext[1]{MSSF construction time is included in Matrix construction time (a).}
\end{table*}

\subsection{Large-scale simulations\label{sec:CPUtime}}

In this section, we investigate the computational efficiency of the multi-site method to demonstrate the practicality of future large-scale simulations.
We compare the computational time for one self-consistent-field (SCF) step with primitive PAOs and with MSSFs for two systems, m0 (1544 atoms) and the same system doubled in size, m0double (3088 atoms).
The time for (a) the construction of the overlap and Hamiltonian matrices and (b) the diagonalization of the Hamiltonian in one SCF step are summarized in Table~\ref{computime}.
We use the subroutine PZHEGVX in ScaLAPACK~\cite{blackford1997} for diagonalisation, and
all of the calculations are done on the supercomputer SGI ICE X (Intel Xeon E5-2680V3 (12 cores, 2.5 GHz)$\times$2 and 128 GB memory per node) at NIMS.

First, we compare the times for m0 moir\'{e} structure with the PAOs (\verb|#1| in the table) and the MSSFs (\verb|#2|). The number of the local orbitals, that is the dimension of the Hamiltonian, with the MSSFs, 8112, is almost 3.4 times smaller than that of the primitive PAOs, 27192. Because of the cubic scaling, the computational time for the diagonalization should ideally be 37.7 times smaller and is found to be 33.8 times smaller. In this MSSF calculation, the computational time to construct the MSSFs is much longer than the time of the Hamiltonian diagonalization. Note that the time to construct MSSFs is large because we use large cutoff region (16 bohr) for the MSSFs in the present calculations. Nevertheless, due to the significant reduction of the computational time in the diagonalization, the total time with the MSSFs is about three times smaller than that with the PAOs.

Next, we turn to the computational times for the larger m0double system.
In the calculation with the PAOs (\verb|#3|), we had to use fewer MPI processes (on more nodes) than in \verb|#1| for optimal memory access. The comparison between \verb|#3| and the calculation with the MSSFs (\verb|#4|) shows that the MSSF method can reduce the diagonalization time dramatically. Although we need additional time to construct the MSSFs, the total computational time is reduced from 37803.5 seconds (more than 10 hours) to 2156.3 seconds (about 0.6 hour)--a speed-up of nearly 20.

We also compare the computational times with the MSSFs for m0 (\verb|#2|) and m0double (\verb|#5|) systems, using the same number of computer nodes and MPI processes. Since the computational time to construct the MSSFs should be linear with system size, the time should be doubled when we double the system size. Table~\ref{computime} shows that it is about 1.8 times larger. The time for diagonalization for m0double should be 8 times larger as that for m0, and it is found to be 6.7 times.

We can investigate the parallel efficiency by comparing the calculations \verb|#6| with 864 MPI processes and \verb|#5| with 432 MPI processes. The data shows that the construction of the matrix elements with MSSFs (a) scales quite well, i.e., the ratio \verb|#5/#6| should be 2 and actually 1.8 in the present calculation. Although the parallel efficiency for the diagonalization (b) is not as good as in (a), the computational time for one SCF step in \verb|#6| is less than 10 minutes and is 1.7 times smaller than that of \verb|#5|. With this computation time, it would be perfectly possible to perform a full DFT study of such large systems, containing more than 3000 atoms.

\section{Conclusions\label{sec:Conclusion}}

In this work, we have presented recent calculations on a graphene-rhodium interface using the the localized orbitals implemented in the \textsc{Conquest} code. We employed both standard basis sets based on PAOs and a multi-site projection which allow us to keep the accuracy level of the larger basis set, while using a small support basis, giving significant savings on computational time and memory requirements. The good agreement between these new results and previous plane-wave calculations shows that the \textsc{Conquest} code is a promising tool for large-scale systems containing thousands of atoms, even for complex systems. The multi-site approach shown here lies between full, primitive basis set calculations for small systems (up to a few hundred atoms) and the linear scaling approach for large systems (over 10,000 atoms). This approach, that can be implemented in other localized orbitals codes, significantly extends the size of system that can be addressed with DFT calculations on modest computational resources, without using linear scaling approaches.

Although \textsc{Conquest} has already shown its potential for many, relatively simple, benchmark systems, our results reported in this work are the first example in which the interaction between delocalized charge states, graphene $\pi$-bands, and a metallic substrate has been successfully addressed with this methodology.  We have demonstrated how it is possible to simulate very large systems, containing over 3000 atoms, using the contracted multi-site basis set. With this methodology the accuracy to resolve subtle details of the larger basis sets is preserved. In the case of graphene-metal interfaces, this includes the correct calculation of adsorption energies, corrugations, adsorption distances, and electronic structure.

%%--------------------------------------------------------------%%
%% ACKNOWLEDGMENTS
%%--------------------------------------------------------------%%

\begin{acknowledgments}
We thank the financial support of the Spanish MINECO (projects MAT2014-54484-P and MAT2017-83273-R). CRM is grateful to FPI-UAM graduate scholarship and NIMS Internship programs and Fundaci\'on Universia for financial support. Calculations in Sec.~\ref{sec:CPUtime} were performed on the Numerical Materials Simulator at NIMS. AN and TM thank the support by JSPS KAKEHI Grant Number 17H05224 and New Energy and Industrial Technology Development Organization of Japan (NEDO) Grant (P16010).
\end{acknowledgments}

%%--------------------------------------------------------------%%
%% BIBLIOGRAPHY
%%--------------------------------------------------------------%%

%\nocite{*}
\bibliography{biblio}% Produces the bibliography via BibTeX.

\end{document}